\documentclass[%
 reprint,%
 amssymb, amsmath,%
 aip,cha,%
]{revtex4-1}

\usepackage{amsmath}

\if x\pdfoutput\undefined  
\usepackage[dvips]{graphicx}  
\else  
\usepackage[pdftex]{graphicx}
\fi  
\graphicspath{{./},{./graphics/}}

\usepackage{natbib}

\usepackage{docs}%
\usepackage{bm}%
\usepackage[colorlinks=true,linkcolor=blue]{hyperref}%
\expandafter\ifx\csname package@font\endcsname\relax\else
 \expandafter\expandafter
 \expandafter\usepackage
 \expandafter\expandafter
 \expandafter{\csname package@font\endcsname}%
\fi
\begin{document}

\title{\bf On the relation between plasticity, friction, and geometry}

\author{S. Barbot}

\email{sbarbot@ntu.edu.sg}
\affiliation{Earth Observatory of Singapore, Nanyang Technological University}%


\begin{abstract}
Plasticity refers to thermodynamically irreversible deformation associated with a change of configuration of materials. Friction is a phenomenological law that describes the forces resisting sliding between two solids or across an embedded dislocation. These two types of constitutive behaviors explain the deformation of a wide range of engineered and natural materials. Yet, they are typically described with distinct physical laws that cloud their inherent connexion. Here, I introduce a multiplicative form of kinematic friction that closely resembles the power-law flow of viscoplastic materials and that regularizes the constitutive behavior at vanishing velocity, with important implications for rupture dynamics. Using a tensor-valued state variable that describes the degree of localization, I describe a constitutive framework compatible with viscoplastic theories that captures the continuum between distributed and localized deformation and for which the frictional response emerges when the deformed region collapses from three to two dimensions. 
\end{abstract}

\maketitle

\section{Introduction}

\begin{figure}
\centering
\includegraphics[width=0.5\textwidth]{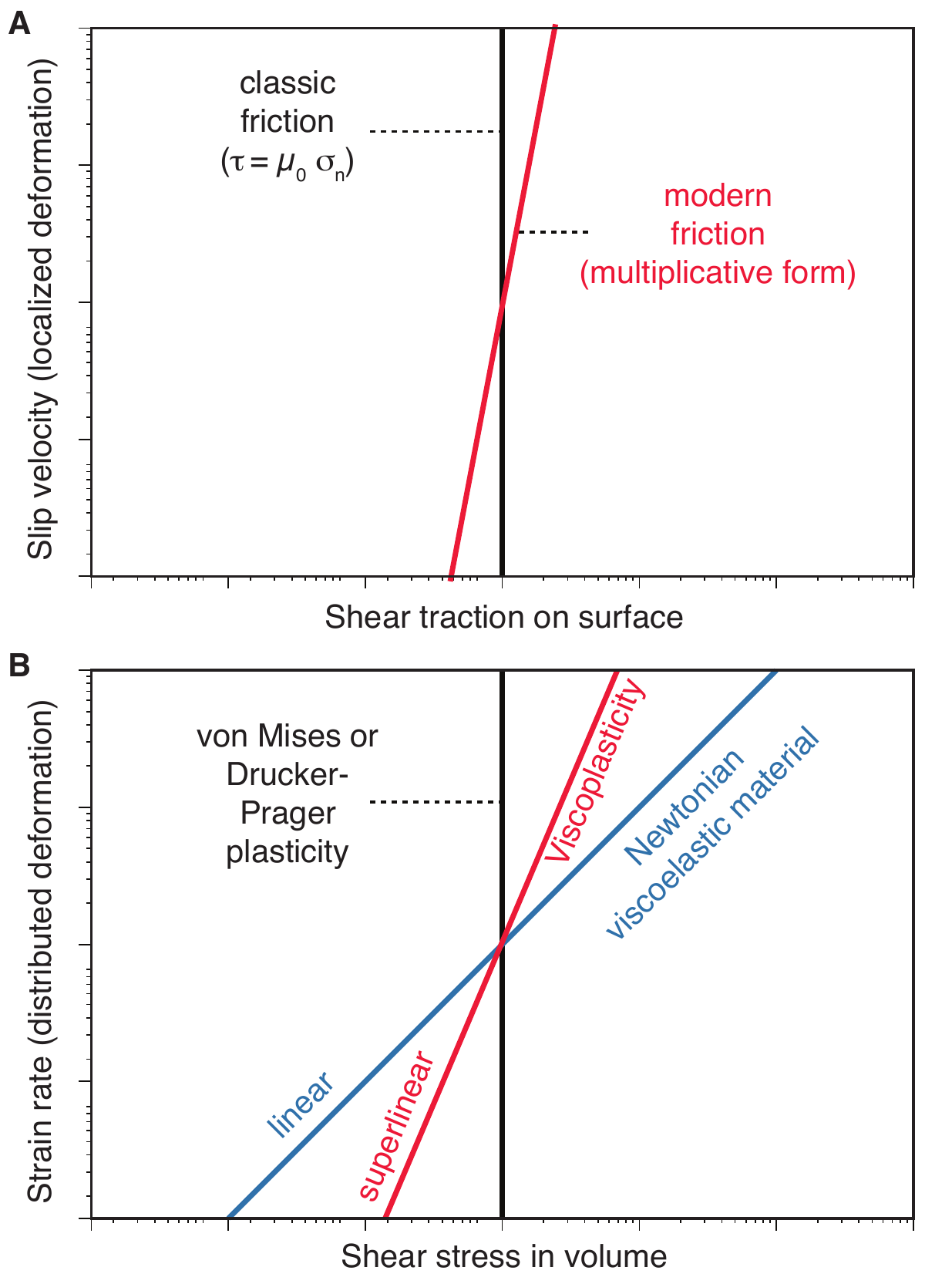}
\caption{Displacement fields in two-dimensional models of deformation due to increasingly localized deformation. We consider an embedded dislocation buried at 20\,km depth, 20\,km long in the down-dip direction, and dipping 45$^\circ$ from the vertical. The left panels are for a screw dislocation and the right ones for an edge dislocation. The top, middle and bottom panels are for shear zones of 10\,km, 5\,km, and 200\,m, respectively. The contour lines are every 20\,cm.}
\label{fig:classic-modern-views}
\end{figure}

The deformation of solid media under stress is complex, often nonlinear, and involves a wide range of microphysical processes. Anelastic deformation represents the thermodynamically irreversible deformation of materials, implying that under this type of deformation materials do not recover their initial configuration if the applied stress is removed. Often, deformation includes both elastic (reversible) and anelastic components. Plastic flow refers to the distributed anelastic deformation of materials under stress and is observed in most engineered and natural materials, including metals, soils, and rocks. Friction describes the resistance to sliding of solid surfaces in contact. Being confined at a solid interface, friction represents a case of localized deformation.

The rheology of plastic flow has been classically described by the von Mises model\cite{huber1904,mises1913} or its variants\cite{drucker+prager52} that assumes no plastic flow when the stress is below a yield surface. Friction has also be described mathematically as a yield surface, implying that sliding may not occur unless the shear forces reach a critical value that depends on normal stress. These results were found experimentally by Da~Vinci, re-discovered by Amontons\cite{amontons1699}, and confirmed by Coulomb\cite{coulomb1785}. These models afford a simple description of the phenomenology but they do not describe how fast the medium flows or breaks when placed under sufficient stress nor the complex dynamics or instabilities that can develop near the yield surface\cite{scholz90a}. 

The limitations of the classic formulations have been remedied by the development of viscoplasticity\cite{andrade1910,norton1929} and the Dieterich-Ruina rate-and-state form of kinematic friction\cite{dieterich78,ruina83}. In the classic view of plastic flow and friction, anelastic deformation occurs only at the yield surface and the stress cannot exceed the yield surface (Fig.~\ref{fig:classic-modern-views}). In the modern view, the yield surface represents a reference point under which deformation is very slow and above which deformation is very fast. The concept of a yield surface is only useful if the anelastic  deformation rate is superlinear with stress. The weak sensitivity of the frictional resistance to sliding velocity noted by Coulomb is captured by the logarithmic sensitivity to velocity in the modern formulation of rate-and-state friction (a wide range of slip velocity occurs for a narrow range of effective stress). For plastic flow, the strength is often rate dependent, taking a linear form for viscoelastic materials and often a power-law form for plastic materials. However, there is a continuum between linear, power-law, and exponential stress-strain-rate relationships embodied by processes such as diffusion creep, dislocation creep\cite{karato+jung03}, and Peierls creep\cite{peierls1940,nabarro1947}. Like friction, plastic flow may also necessitate state variables for phenomena associated with transient creep\cite{masuti+16} and work-hardening in general\cite{nemat-nasser04}. The viscoplasticity and rate-and-state friction theories have allowed modeling multiple cycles of ruptures with complex deformation histories\cite{lambert+barbot16,allison+dunham17}.

As noted by others\cite{michalowski+mroz78,curnier84}, the similitudes in the parallel development of the theories of plasticity and friction are striking. The mechanics of friction and plasticity are also deeply related. For example the micromechanics of plasticity can be explained as the motion of internal dislocations\cite{volterra1907,orowan1934} and the rate- and state-dependence of friction can be explained by the plastic behavior of the contact asperities regulating the real area of contact\cite{bowden+tabor50,rabinowicz58,bowden+tabor64}. In many cases, the frictional interface consists in a finite-width shear zone made of granular, damaged, or semi-brittle material and the assumption of a simple frictional behavior is for mathematical or modeling convenience only\cite{steketee58a,barbot+fialko10a}. This suggests that these theories could be combined in a more general, unifying framework. 

Plasticity and friction represent the end-members for distributed and localized deformation, respectively. However, they cannot capture the continuum between localized and distributed deformation. This can be particularly important when describing the inception of new frictional surfaces by the accumulation of damage or increased localization. During this process, the material first deforms plastically until a frictional interface has matured. When this point is reached, a new surface has emerged and friction can become the dominant deformation mechanism.

In geophysics, there is an impetus to combine long-term and short-term processes to allow faults to emerge from tectonic processes and resolve rupture dynamics on these new structures\cite{herrendorfer+15} as important feedbacks operate at these different time scales. That dynamic ruptures on frictional interfaces and plastic deformation operate at different time scales represents a significant numerical challenge. But there is no convenient theoretical framework to describe friction and plasticity in a unified way. 

Towards that objective, I highlight two important results. In Section~\ref{sec:kinematic}, I show that the kinematics of distributed deformation and localized deformation can be captured by a single mathematical expression\cite{barbot+17}. This highlights a continuum of kinematic descriptions between distributed and localized deformation. In Section~\ref{sec:rheology}, I show that the modern rheological laws that describe plastic flow have a striking similarity with the ones describing frictional dynamics, despite some subtle but important differences. To do so, I introduce the multiplicative form of rate-and-state friction. In Section~\ref{sec:unified}, I conjecture a constitutive framework that reduces to viscoplasticity in some situations and to kinematic friction framework in others. The model predicts that frictional behavior is an end-member rheological response due to the collapse of deformation region from three to two dimensions. 

In this paper, I do not discuss a new microphysical model of plasticity or friction. Such important models shed light on the constitutive equations of anelasticity and can be found in the work of various authors\cite{dieterich+kilgore94,sleep05,niemeijer+spiers07,putelat+11,perfettini+molinari17,molinari+perfettini17}. Instead, I discuss the relation between the macroscopic representations of plasticity and friction in a continuum that are directly incorporated in analytic or numerical models of time-dependent deformation.

\section{Kinematic continuum between distributed and localized deformation}\label{sec:kinematic}

In this Section, I consider the kinematics of deformation of plastic and brittle materials and how anelastic deformation induces strain in the surrounding medium by elastic coupling. The static or quasi-static equilibrium attained in the presence of anelastic deformation can be modeled using transformation strain\cite{nemat-nasser04,nemat-nasser+hori99}, which can be due to thermal stress, multi-phase flow, or plastic deformation\cite{barbot+fialko10a}. Steketee\cite{steketee58a,steketee58b} showed that the Schmidt tensor, a particular form of transformation strain, can capture the motion of dislocations. However, because of the fundamental difference in nature between distributed and localized deformation, these cases were historically treated distinctly. 

\begin{figure*}[p]
\centering
\includegraphics[width=0.85\textwidth]{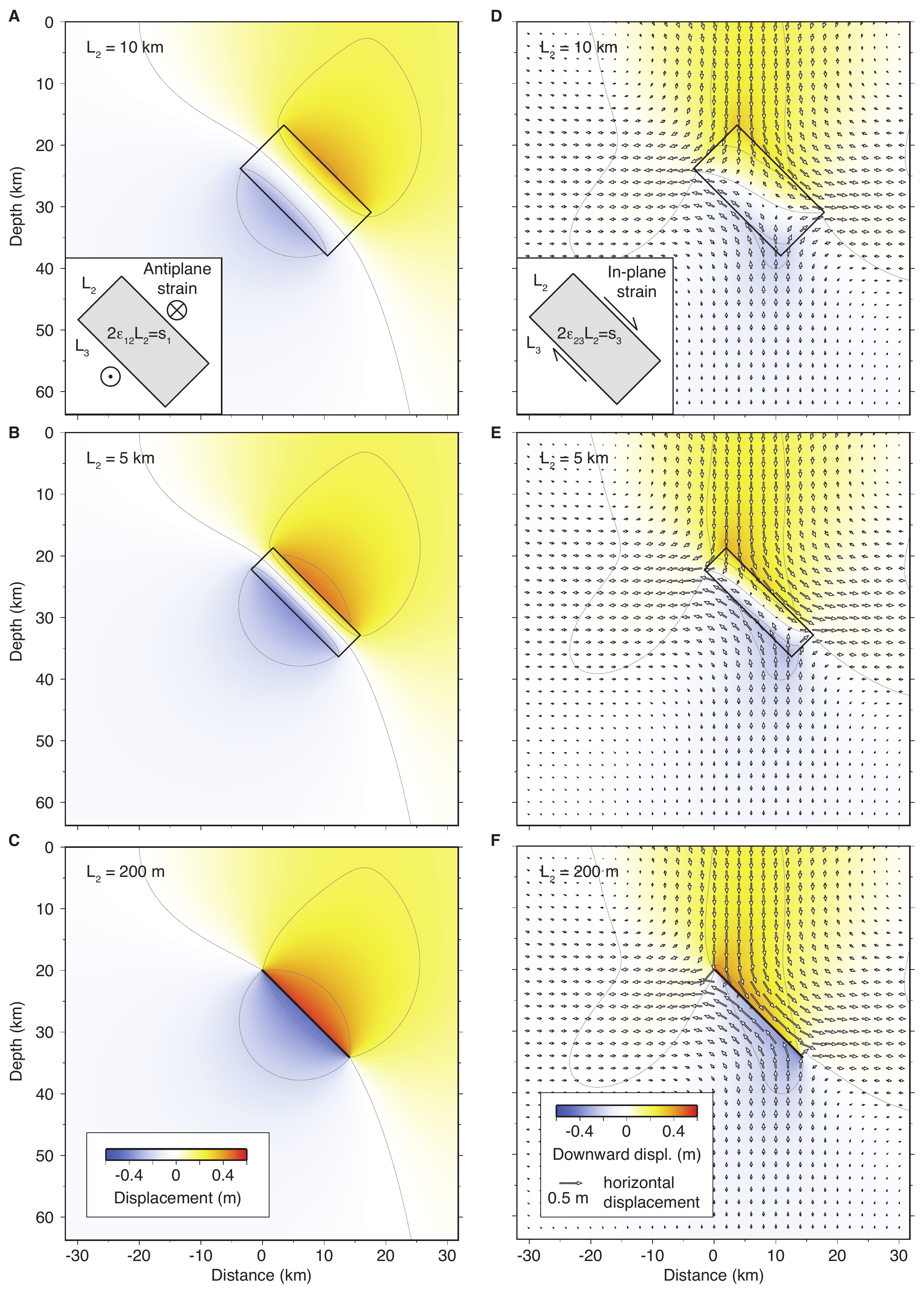}
\caption{Displacement fields due to increasingly localized deformation. We consider the plastic deformation of a cuboid representative volume element (rectangle) buried at 20\,km depth, 20\,km long in the down-dip direction, and dipping 45$^\circ$ from the vertical. The left panels are for out-of-plane plastic strain and the right ones for in-plane plastic strain. The top, middle and bottom panels are for shear zones of 10\,km, 5\,km, and 200\,m, respectively. The contour lines are every 20\,cm. With increasing localization of plastic deformation, the deformation converges to the cases of a screw dislocation and an edge dislocation for the left and right panels, respectively.}
\label{fig:kinematic-distributed-localized}
\end{figure*}

Here, I show how various degrees of localization can be described within a single framework. To do so, I consider the plastic deformation of a shear zone confined in a larger continuum and subjected to the transformation strain components $\epsilon_{ij}$. The region that deforms anelastically takes the form of a cuboid with dimension $L_1$, $L_2$, and $L_3$. For convenience, let us imagine a reference system aligned with the cuboid so that the subscript indices match those of the basis vectors. The resulting deformation can be evaluated by solving the governing equation for elasticity using equivalent body forces to incorporate the plastic strain\cite{barbot+fialko10a,barbot+17,barbot18}. For illustration purposes, I consider a half space with a free surface and the cases of in-plane and anti-plane deformation.

I subject the shear zone to increasing plastic strain in a cuboid of decreasing width $L_2$. For the sake of simplicity, I adopt a two-dimensional model where $L_1=\infty$ and I choose a grid size of 250\,m for all calculations, which also represents the dimension of a representative volume element. I reduce the plastic strain to the nontrivial component $\epsilon_{23}$ for the case of in-plane strain and to the nontrivial component $\epsilon_{12}$ for the case of anti-plane strain. The cumulative displacement across the shear zone due to plasticity alone, i.e., irreversible, is
\begin{equation}
s_3=2\epsilon_{23} L_2
\end{equation}
for in-plane strain and
\begin{equation}
s_1=2\epsilon_{12} L_2
\end{equation}
for anti-plane strain. In these simulations, I keep $s_1$ and $s_3$ constant and I reduce $L_2$ incrementally. The resulting deformations are shown in Fig.~\ref{fig:kinematic-distributed-localized}. When the width of the shear zone is lower than the grid size, the shear zone virtually collapses from three to two dimensions, i.e., from a volume to a surface. The solutions converge towards the one for an edge dislocation for in-plane strain and towards the one for a screw dislocation for anti-plane strain. Using in-plane strain as an example, the limit case of localized deformation is obtained for
\begin{equation}
\epsilon_{23}=\lim_{L_2\rightarrow 0}\frac{s_3}{2L_2}~.
\end{equation}
That is, the motion of dislocation is associated with infinite strain. For brittle deformation, plastic strain is infinite, the support is zero, but the product $s_3=2\epsilon_{23} L_2$ is finite and the cumulative plastic displacement is called slip\cite{michalowski+mroz78}. 

These results indicate that brittle deformation can be thought of as an end-member of plasticity for infinite strain over a zero volume, at least in terms of the kinematics of deformation. These results have been exploited to simulate fault slip in the framework of transformation strain using numerical methods\cite{steketee58a,barbot+fialko10a} and to derive analytic expressions for the deformation of elasto-plastic media for various degrees of localization, from distributed to localized\cite{barbot+17,barbot18}. In reality, even the dislocations associated with the motion of lattice defects in crystal structures have a finite width commensurate to the atomic scale ($L_2\gtrsim1\,$\r{A}), so plastic strain is always bounded. 

Unified kinematic models of anelastic deformation from distributed to localized have been explored\cite{barbot+fialko10a,barbot+17} and form a motivation to explore the possibility of a unification of the constitutive laws.

\section{The multiplicative form of kinematic friction and its relation to viscoplasticity}\label{sec:rheology}

\begin{figure}
\centering
\includegraphics[width=0.5\textwidth]{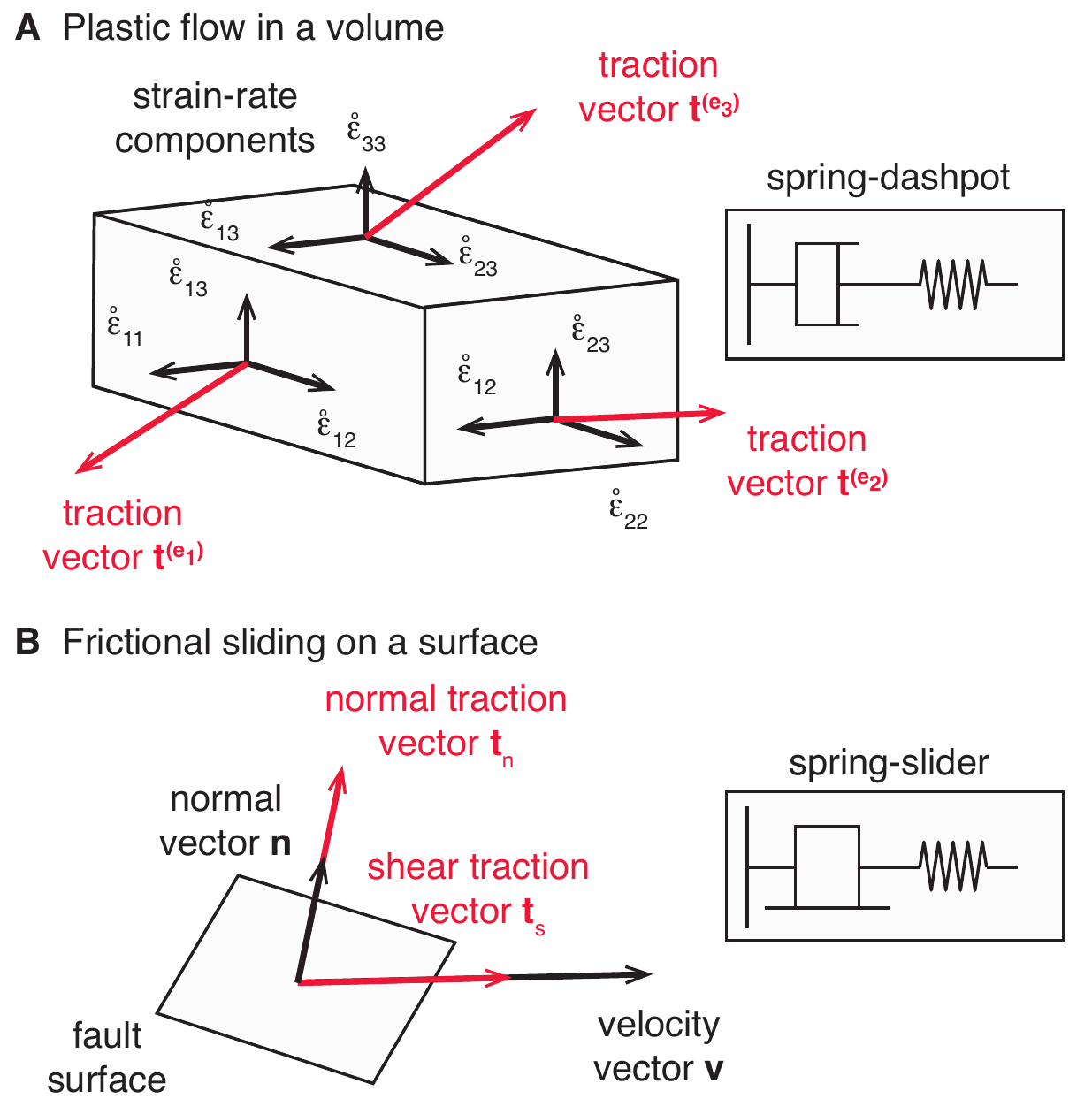}
\caption{Representative elements for viscoplasticity and kinematic friction. A) The dynamics of viscoplastic flow is represented by a stress-strain-rate constitutive relationship, the simplest one being a spring-dashpot model. B) The dynamics of frictional sliding relies on a velocity-stress constitutive relationship that incorporates the orientation of the representative surface element. The simplest model is a spring-slider.}
\label{fig:fault-cube-stress-traction}
\end{figure}

Viscoplasticity describes the rate-dependent deformation of plastic materials under stress\cite{chaboche08}. The model describes the response of a representative volume element (Fig.~\ref{fig:fault-cube-stress-traction}a). The simplest model for viscoplasticity is a combination of a spring and a (potentially nonlinear) dashpot. Broadly speaking, the rate of plastic strain can be described as
\begin{equation}
\begin{aligned}
\dot{\boldsymbol{\epsilon}}&=\dot{\boldsymbol{\epsilon}}(\boldsymbol{\sigma},\boldsymbol{\epsilon},\psi_i)~,
\end{aligned}
\end{equation}
where $\dot{\boldsymbol{\epsilon}}$ is the plastic strain rate tensor, $\boldsymbol{\sigma}$ is the stress tensor, and the $\psi_i$ are state variables following the evolution laws
\begin{equation}
\begin{aligned}
\dot{\psi_i}&=\dot{\psi_i}(\dot{\boldsymbol{\epsilon}},\boldsymbol{\sigma},\psi_i)~.
\end{aligned}
\end{equation}
The inclusion of state variables and the cumulative strain in the constitutive behavior allows for the transient and work-hardening effects and makes the rate of deformation non-conservative, i.e., dissipative and path dependent.

Modern friction describes the rate of sliding across a representative surface element upon application of shear and normal tractions\cite{dieterich78,ruina83}. The simplest constitutive model is to combine a spring and a slider (Fig.~\ref{fig:fault-cube-stress-traction}b). Rate-and-state friction is a phenomenological law that describes the frictional resistance to sliding including healing and weakening. This framework has some known shortcomings\cite{nakatani01,bayart+06,segall10}, but it has nonetheless been successful at explaining many dynamic mechanical systems over a wide range of materials and boundary conditions\cite{tse+rice86,lapusta+00,carlson+batista96,lemaitre02,aranson+02,urbakh+04,lapusta+barbot12,barbot+12,meleveedu+barbot16,kaneko+16}. A friction law can be written as follows
\begin{equation}\label{eqn:friction-constitutive}
\textbf{v}=\textbf{v}(\boldsymbol{\sigma},\theta_i\,;\textbf{n})~,
\end{equation}
where $\textbf{v}$ is the sliding velocity, the $\theta_i$ are state variables with the evolution laws
\begin{equation}
\begin{aligned}
\dot{\theta_i}&=\dot{\theta_i}(\textbf{v},\boldsymbol{\sigma},\theta_i\,;\textbf{n})~,
\end{aligned}
\end{equation}
and $\textbf{n}$ is the local normal direction of the surface where frictional sliding occurs. The notable difference between plasticity and friction is the role of geometry. The constitutive equation for viscoplasticity applies everywhere in the medium. In constrast, the traction and the slip vector cannot be established without a proper representation of the surface in terms of its orientation and position. Indeed, frictional sliding may be associated with the transformation strain
\begin{equation}
\textbf{R}=\frac{1}{2}\left(\textbf{v}\otimes\textbf{n}+\textbf{n}\otimes\textbf{v}\right)\,\delta\left(\textbf{x}\cdot\textbf{n}\right)
\end{equation}
that explicitly incorporates the normal vector of the frictional surface. Any constitutive law for anelastic deformation that does not explicitly include geometry cannot fully represent friction.

\subsection{The multiplicative and additive forms of kinematic friction}

\begin{figure}
\centering
\includegraphics[width=0.5\textwidth]{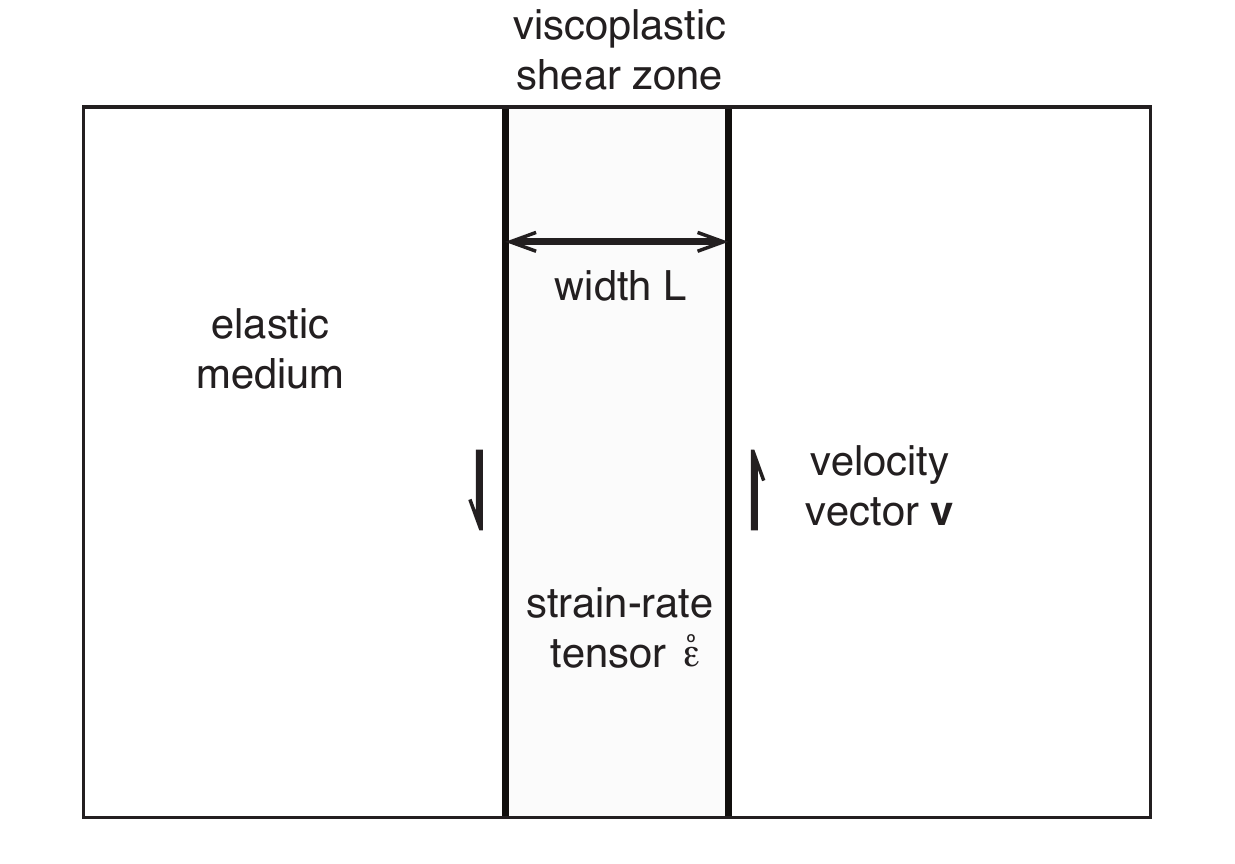}
\caption{A finite shear zone model for frictional sliding. The relative velocity across frictional interface is accommodated by plastic strain within the shear zone.}
\label{fig:shear-zone-plasticity-friction}
\end{figure}

With this important difference in mind, I now describe some similarities in form between the constitutive laws for viscoplasticity and kinematic friction. I consider a thought experiment where the frictional interface is modeled by a  narrow shear zone of width $L$ undergoing plastic deformation (Fig.~\ref{fig:shear-zone-plasticity-friction}) with the following rheology
\begin{equation}\label{eqn:equivalence-starting-point}
\frac{\dot{\epsilon}}{\dot{\epsilon}_0}= \left(\frac{\tau}{\mu_0\bar{\sigma}}\right)^n\left(\frac{d_0}{d}\right)^m~,
\end{equation}
where $\dot{\epsilon}_0$ is a reference strain rate, $\tau$ is the norm of the shear stress in the direction of sliding, $d$ is the average grain size in a gouge layer, $d_0$ is a reference value, and $\mu_0\bar{\sigma}$ is a reference stress. I assume that grain size is a state variable that varies with
\begin{equation}\label{eqn:power-law-evolution}
\dot{d}=V_0 - 2\dot{\epsilon}d~,
\end{equation}
where the first term in the right-hand side gives rise to healing and the second term accounts for dynamic recrystallization\cite{hall+parmentier03}. Classically, viscoplastic rheologies are formulated with a different nomenclature, but I make this choice of parameterization for the sake of the argument. The power exponent is assumed in the range $n=50-100$, corresponding to low-temperature creep. The grain size exponent is assumed in the range $m\in[0,3]$, the bounds corresponding to negligible and significant dependence on grain size, respectively.

The above formulation can be shown to be mathematically equivalent to rate-and-state friction with the aging law with the following few assumptions. First, to ease the comparison, I turn the strain rate into a velocity using $V=2L\dot{\epsilon}$ and $V_0=2L\dot{\epsilon}_0$. Using the state variable $d=\theta V_0$, with $d_0=\theta_0 V_0$, I obtain
\begin{equation}\label{eqn:power-law-friction-velocity}
\frac{V}{V_0}=\left(\frac{\tau}{\mu_0\bar{\sigma}}\right)^n\left(\frac{\theta_0}{\theta}\right)^m~.
\end{equation}
Then, I recast the stress-strain-rate relationship in terms of frictional resistance
\begin{equation}\label{eqn:power-law-friction}
\tau=\mu_0\bar{\sigma}\left(\frac{V}{V_0}\right)^{1/n}\left(\frac{\theta}{\theta_0}\right)^{m/n}~.
\end{equation}
As the sensitivity of stress to velocity is only weak, I consider a Taylor series expansion using $x^y=e^{y\ln x}=1+y\ln x+\mathcal{O}(\ln^2 x)$ around $x=1$, expanding the relationship (\ref{eqn:power-law-friction}) accordingly around $V=V_0$. Defining friction as $\mu=\tau/\bar{\sigma}$ and keeping only the linear terms, I obtain
\begin{equation}\label{eqn:power-law-additive}
\mu=\mu_0\left[1+\frac{1}{n}\ln\frac{V}{V_0}+\frac{m}{n}\ln\frac{\theta}{\theta_0}\right]~.
\end{equation}
The correspondence with the classic nomenclature of rate-and-state friction is immediate if we note $a=\mu_0/n$, $b=ma$, and $d_0=L$. To go further let us consider the evolution law~(\ref{eqn:power-law-evolution}). Turning the strain rate into a velocity, I obtain
\begin{equation}
\begin{aligned}
\dot{d}&=\dot{\theta} V_0= V_0 - \frac{V}{L}\theta V_0~,
\end{aligned}
\end{equation}
or, simply,
\begin{equation}\label{eqn:aginglaw}
\begin{aligned}
\dot{\theta}& = 1 - \frac{V\theta}{L}~,
\end{aligned}
\end{equation}
which is the aging law of rate-and-state friction\cite{ruina83}. In this formulation the state variable is interpreted as the average grain size scaled by a reference sliding velocity and the reference grain size coincides with the width of the gouge layer, which also matches with the characteristic weakening distance. The direct effect coefficient $a=\partial \mu / \partial \ln V$ is related to the reciprocal of the power exponent, consistent with values of $a$ of the order of $10^{-2}$. The $b$ parameter is controlled by the dependence to grain size. 

I note that there is no laboratory experiment that describes a combination of power-law flow with a high stress exponent with a grain-size dependence such as in (\ref{eqn:equivalence-starting-point}), although it complies to the case of dislocation glide\cite{warren+hirth06} with $n=2$. Nonetheless, the plastic model for friction encapsulated in equations (\ref{eqn:equivalence-starting-point}) and (\ref{eqn:power-law-evolution}) is validated a posteriori by a mathematical equivalence with rate-and-state friction, an empirical law tuned to a number of laboratory experiments. Admittedly, the proof is only mathematical, and therefore the evolution law (\ref{eqn:power-law-evolution}) may not represent grain size but any other physical property that varies with the same rate. 

I call (\ref{eqn:power-law-friction-velocity}) and (\ref{eqn:power-law-friction}) the multiplicative or power-law form of rate-and-state friction, and the Dieterich-Ruina formulation (\ref{eqn:power-law-additive}) the additive form of rate-and-state friction. The multiplicative form is more appealing than the additive one because the latter is ill-posed for vanishing velocities. Indeed, (\ref{eqn:power-law-additive}) incorrectly predicts a negative norm of the traction vector for sufficiently small velocities and an infinite value for a vanishing velocity. The viscoplastic model of frictional resistance also brings new insight. Indeed, some laboratory measurements support that the critical slip distance correlates with the width of the frictional interface\cite{marone+kilgore93,marone98a}. 

I emphasize that (\ref{eqn:power-law-friction-velocity}) is not formally a viscoplastic law because the dynamic variables include the shear and normal traction instead of the deviatoric stress and the pressure. That is, the geometric layout of the shear zone has been incorporated in (\ref{eqn:power-law-friction-velocity}) to conform to the general case~(\ref{eqn:friction-constitutive}). However, the similarity between (\ref{eqn:power-law-friction-velocity}) and the power-law form of viscoplasticity is striking and hints at the existence of a constitutive relationship that incorporates plasticity and friction as end-members. 

\subsection{Comparison with laboratory data}

\begin{figure*}[t]
\centering
\includegraphics[width=1.2\columnwidth]{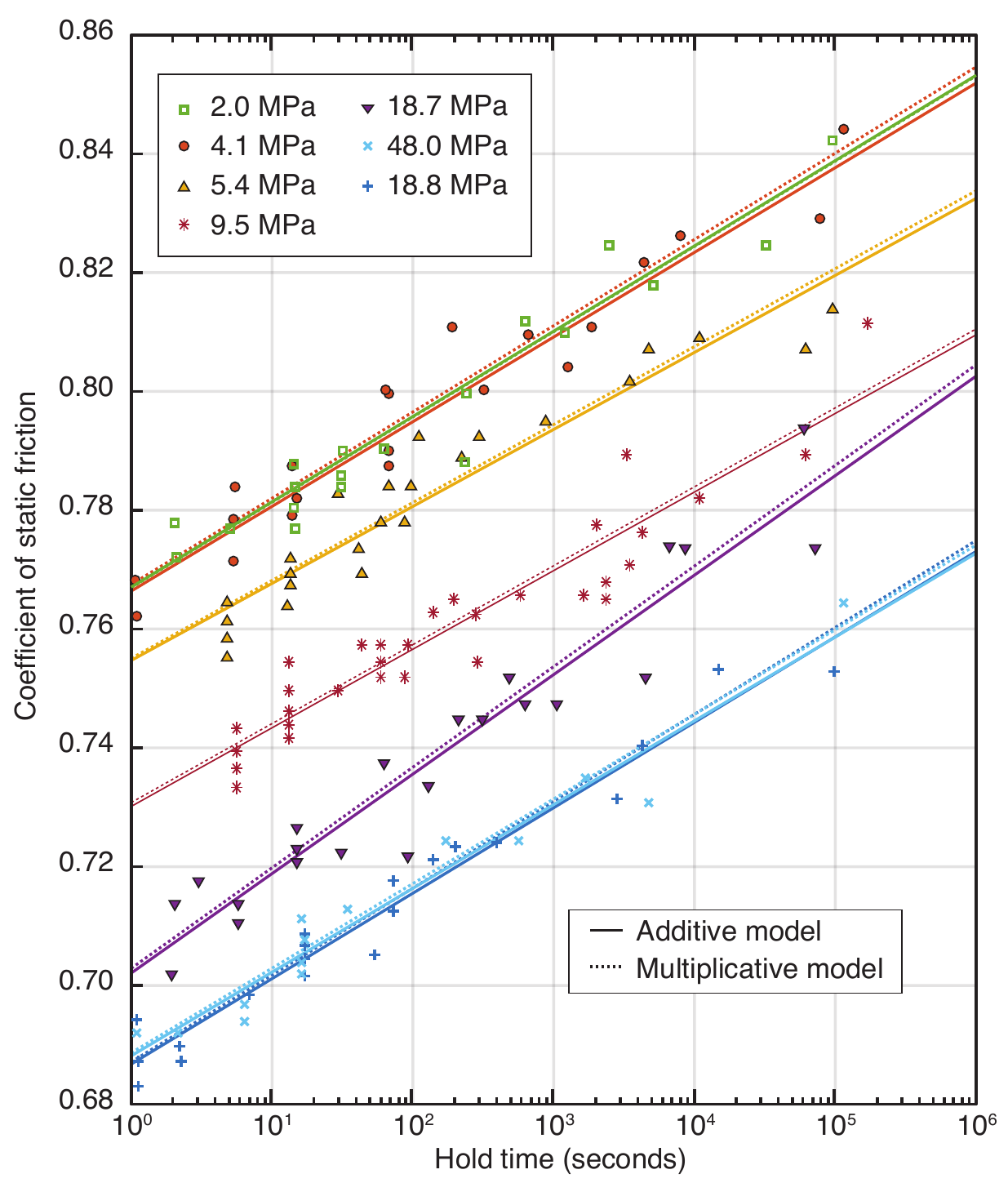}
\caption{Healing and the additive and multiplicative forms of rate-and-state friction. The static coefficient of friction data for quartz sandstone from\cite{dieterich72} for hold tests. The hold time corresponds to the duration of stick. The data are reduced with the additive model (\ref{eqn:power-law-additive}) (solid lines) and the multiplicative model (\ref{eqn:power-law-friction}) (dashed profiles). Both models explain the data well within their scatter.}
\label{fig:comparison_fit_multiplicative_additive_dieterich72}
\end{figure*}

The choice of a constitutive framework may have important implications on the predictions of failure modes, from laboratory samples to large-scale geodynamics. Therefore, one should proceed carefully when introducing a different friction law. The additive form (\ref{eqn:power-law-additive}) of rate-and-state friction was derived empirically by Dieterich\cite{dieterich78} and Ruina\cite{ruina83} based partially on hold experiments that allowed re-strengthening at zero slip rate\cite{dieterich72}. Here, I revisit the data collected by Dieterich\cite{dieterich72} to assess the merit of using (\ref{eqn:power-law-friction-velocity}) to explain these observations. In Fig.~\ref{fig:comparison_fit_multiplicative_additive_dieterich72}, I show the static coefficient of friction of five laboratory experiments conducted at different normal stresses for a wide range of hold times. While different, the predictions of the additive (solid profiles) and multiplicative (dashed profiles) forms of rate-and-state friction are both well within the data scatter and have virtually equal merit in reducing the laboratory data.

The apparent linear scaling of the static coefficient of friction with the logarithm of hold time ($\mu_0\sim\ln t$) motivated Dieterich for his now classic formulation. The multiplicative form of rate-and-state friction (\ref{eqn:power-law-friction-velocity}) implies a different scaling, namely a linear dependence between the logarithm of the coefficient of friction with the logarithm of hold time ($\ln\mu_0\sim\ln t$). Some broader considerations justify the latter as a better choice. 

Tarantola\cite{tarantola06} showed that the natural distance between two physical quantities $\tau_1$ and $\tau_2$ that cannot be negative, which he refers to as \textit{Jeffreys quantities}, is $\left|\ln\tau_1/\tau_2\right|$. This metric is insensitive to the choice of physical unit, such that $\left|\ln\tau_1/\tau_2\right|$ is the same whether $\tau_1$ and $\tau_2$ are expressed in Pa or MPa. As Jeffreys quantities live in the interval $(0,\infty)$, their associated random variable cannot be normally distributed and their non-informative prior cannot be uniformly distributed, both precluding using simple least squares for parameter estimation. The yield stress and the amplitude of the traction vector, which are equated through the consistency relation $\tau=\mu(V,\theta)\bar{\sigma}$, are Jeffrey parameters. Therefore, it is preferable to consider the metric $|\ln(\mu_1/\mu_2)|$ rather than $|\mu_2-\mu_1|$ to describe the difference between any two friction coefficients. The same argument applies for the velocity dependence $V=V(\tau,\theta)$, as velocity and its conjugate, slowness, are also Jeffrey's parameters. Consequently, a power-law form for the yield stress dependence on the velocity and the state variable is better suited than a logarithmic form.

\subsection{The grain-size evolution of Hall and Parmentier}

Several evolution laws have been proposed in the field of viscoplasticity that may inspire new developments in kinematic friction. Hall and Parmentier\cite{hall+parmentier03} introduce an evolution law for grain size where the thermally activated grain growth reduces the interfacial energy associated with grain boundaries and grain size reduction occurs by sub-grain rotation. They write
\begin{equation}\label{eqn:hall+parmentier03}
\dot{d}=\gamma\,p^{-1}\,d^{1-p}-\lambda \dot{\epsilon}d~,
\end{equation}
where $\gamma$ is temperature dependent and controls the healing rate, $\lambda$ is a non-dimensional parameter controlling the rate of grain size reduction, and $p$ is a non-dimensional material parameter in the range 1 to 4. To reduce the evolution law (\ref{eqn:hall+parmentier03}) to a reference form, I conduct the following change of variable. First, using the geometric setting of Fig.~\ref{fig:shear-zone-plasticity-friction}, I approximate the strain rate by the ratio of the cumulative plastic velocity across the shear zone scaled by the width
\begin{equation}
\begin{aligned}
\dot{\epsilon}&=\frac{V}{2T}~,\\
\dot{\epsilon}_0&=\frac{V_0}{2T}~,
\end{aligned}
\end{equation}
where I used a similar scaling for the reference strain rate. Then, I introduce a new state variable that scales linearly with grain size
\begin{equation}\label{eqn:grain-state}
\theta=d\left(\frac{p}{\gamma}\right)^{1/p}~. \\
\end{equation}
Finally, I introduce a critical distance
\begin{equation}
L=\frac{2T}{\lambda}~.
\end{equation}
With these changes, the coupled constitutive equations (\ref{eqn:equivalence-starting-point}) and (\ref{eqn:hall+parmentier03}) become
\begin{equation}\label{eqn:hall+parmentier03-canonical}
\begin{aligned}
V&=V_0\left(\frac{\tau}{\tau_0}\right)^n\left(\frac{\theta_0}{\theta}\right)^m~,\\
\dot{\theta}&=\theta^{1-p}-\frac{V\theta}{L}~.
\end{aligned}
\end{equation}
This formulation closely resembles the power-law form of rate-and-state friction. Indeed, the grain-size evolution law~(\ref{eqn:hall+parmentier03}) takes the form of the aging law (\ref{eqn:aginglaw}) of rate-and-state friction for $p=1$. However, if the steady-state grain size obeys the piezometric relationship\cite{vwal+93}, then grain-size evolution only leads to velocity-strengthening behavior (e.g., $0\le m \le 1$).

Because the mechanisms of grain-size evolution during plastic flow and the evolution of effective contact area during frictional sliding may be different, the similarities between these evolution laws indicate that multiple mechanisms of material reconfiguration at the microscopic level may take a similar form mathematically and sometimes produce the same macroscopic constitutive behavior.

\subsection{The grain-size evolution of Rozel, Ricard and Bercovici}

Rozel, Ricard and Bercovici\cite{rozel+11} employ basic non-equilibrium thermodynamics to propose a general equation for the mean grain size evolution under the assumption that the whole grain size distribution remains self-similar. They obtain the evolution law of the form
\begin{equation}\label{eqn:rozel-classic}
\dot{d}=\gamma\,p^{-1}\,d^{1-p}-\lambda \frac{d^2}{d_o}\frac{\tau}{\tau_0}\dot{\epsilon}~,
\end{equation}
where the healing term is similar to the formulation of Hall and Parmentier\cite{hall+parmentier03}, but the rate of grain size reduction is proportional to the shear stress. Assuming the geometrical setting of Fig.~\ref{fig:shear-zone-plasticity-friction} and the constitutive equation (\ref{eqn:equivalence-starting-point}), we have $\tau\sim\tau_0$ (recall the weak dependence of frictional strength on sliding velocity). Therefore, the evolution law (\ref{eqn:rozel-classic}) can be simplified to
\begin{equation}\label{eqn:rozel11}
\dot{d}=\gamma\,p^{-1}\,d^{1-p}-\lambda \left(\frac{d}{d_o}\right)^q\dot{\epsilon}d
\end{equation}
in the context of a shear zone, where $q=1$ for the law of Rozel, Ricard and Bercovici\cite{rozel+11} and $q=0$ for the law of Hall and Parmentier\cite{hall+parmentier03}. The weakening term is now similar to the one in (\ref{eqn:hall+parmentier03}), except for a factor $(d/d_0)^q$. To reduce the evolution law (\ref{eqn:rozel11}) to a reference form, I use the state variable (\ref{eqn:grain-state}) and introduce the scaling parameter
\begin{equation}
L=\frac{2T}{\lambda}\,{d_o}^q\left(\frac{p}{\gamma}\right)^{q/p}~.
\end{equation}
With these changes, the coupled constitutive equations (\ref{eqn:power-law-friction}) and (\ref{eqn:rozel11}) become
\begin{equation}\label{eqn:hall+parmentier03-canonical}
\begin{aligned}
V&=V_0\left(\frac{\tau}{\tau_0}\right)^n\left(\frac{\theta_0}{\theta}\right)^m~,\\
\dot{\theta}&=\theta^{1-p}-\frac{V\theta^{1+q}}{L}~.
\end{aligned}
\end{equation}
Because of the choice of parameterization, eq.~(\ref{eqn:hall+parmentier03-canonical}) reduces to the evolution law of Hall and Parmentier\cite{hall+parmentier03} and the aging law of rate-and-state friction for specific values of $p$ and $q$. For example, the formulation simplifies to the aging law of rate-and-state friction for $p=1$ and $q=0$.

These two examples demonstrate that largely different assumptions about the microphysics of the state evolution may lead to similar mathematical expressions for the evolution law. This may explain why the rate-and-state framework has been so successful at capturing deformation in the laboratory for so many different types of materials and stress regimes. 

\subsection{Implications for rupture dynamics}

\begin{figure*}[t]
\centering
\includegraphics[width=1.4\columnwidth]{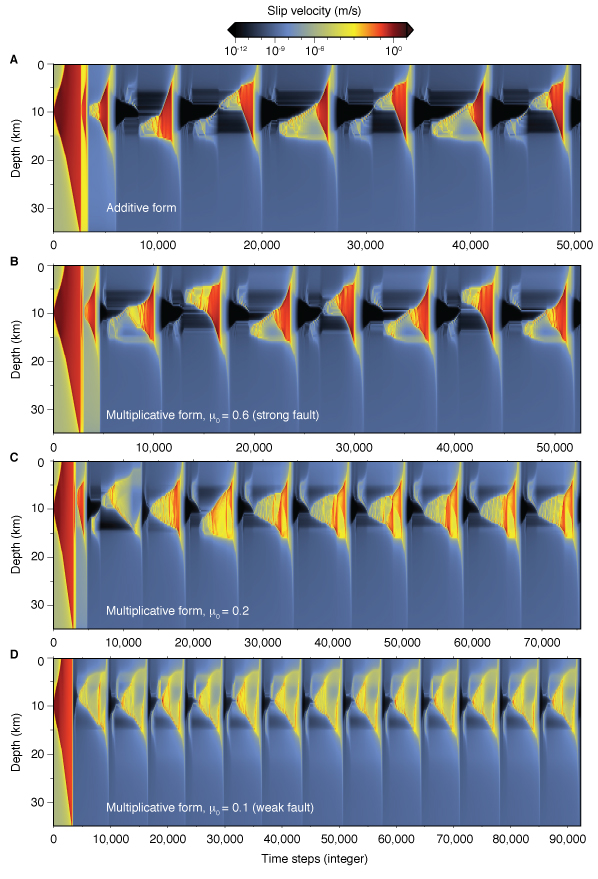}
\caption{Earthquake cycle simulations with the additive (\ref{eqn:power-law-additive}) and multiplicative (\ref{eqn:power-law-friction}) forms of rate-and-state friction. A) Simulations with the additive form, which is insensitive to the value of the static friction coefficient. B, C, D) Simulations with the multiplicative form of rate-and-state friction with a static friction coefficient of $\mu_0=0.6$, 0.2, and 0.1, respectively. Rupture dynamics with the multiplicative form is sensitive to the strength of the fault, which among other characteristics affects the nucleation, the source-time function, and the static and dynamic stress drops of each event.}
\label{fig:comparison_eqcycle_additive_multiplicative}
\end{figure*}

The dynamics of frictional instabilities depends on the details of the constitutive framework. To illustrate the importance of understanding the frictional resistance near zero slip speed, I explore the predictions of the additive and multiplicative forms of rate-and-state friction with numerical models of mode III ruptures in anti-plane strain. Particularly, I show that the rupture modes (the peak velocity, stress drop, stability) strongly depend on the static coefficient of friction when the multiplicative form of rate-and-state friction is assumed.

I consider a vertical frictional interface embedded in an elastic layer with a free surface. Instead of solving for the inertial contribution directly, I use radiation damping as an approximation within the boundary integral method, as in previous work\cite{qqiu+16,lambert+barbot16,goswami+barbot18}. I consider a 35\,km vertical fault that is velocity weakening between 5 and 15\,km depth with $a-b=-4\times 10^{-3}$ and velocity strengthening elsewhere with $a-b=4\times 10^{-3}$, all under a uniform normal stress of $\bar{\sigma}=100\,$MPa. The system is loaded at a uniform rate of $10^{-9}\,$m/s for $2\times 10^{10}\,$s and the initial condition is chosen for steady state at that slip velocity. 

The results shown in Fig.~\ref{fig:comparison_eqcycle_additive_multiplicative} compare dynamic rupture simulations assuming the additive (top panel) and the multiplicative (three bottom panels) forms of rate-and-state friction. For a strong fault with $\mu_0=0.6$, the results from both forms are similar in terms of recurrence times, overall number of instabilities, peak velocities and stress drops. As the static coefficient of friction decreases, the peak velocity and stress drop decrease and the details of the rupture evolution change significantly, except for the first event in the simulation that is strongly influenced by the initial condition. The importance of the static coefficient of friction comes about from the relative amplitude of the yield stress and the stress drop during failure and this effect should be exacerbated by seismic radiations.

The additive form of rate-and-state friction has been interpreted in terms of statistical physics\cite{rice+benzion96,lapusta+00} as a probability for forward jumps of the microasperities forming the real area of contact proportional to $\exp(\tau)$. When backward jumps are taken into account, the probability of forward jumps becomes proportional to $\exp(\tau)-\exp(-\tau)$, leading to a term in $\mathrm{arcsinh}$ instead of $\ln$, as in\cite{rice+benzion96,lapusta+00}
\begin{equation}\label{eqn:lapusta+00}
\tau=a\bar{\sigma}\,\mathrm{arcsinh}\left[\frac{V}{V_0}\,\exp\left(\frac{\mu_0+b\ln(V_0\theta/L)}{a}\right)\right]~.
\end{equation}
The multiplicative form (\ref{eqn:power-law-friction}) assumes the probability of forward jumps of the form $\exp(\alpha\ln\tau)$, i.e., a type of power-law distribution. Further laboratory experiments may be needed to constrain the constitutive behavior at small slip velocity.

\section{A unifying constitutive framework for plastic flow and friction}\label{sec:unified}

\begin{figure}
\centering
\includegraphics[width=0.9\columnwidth]{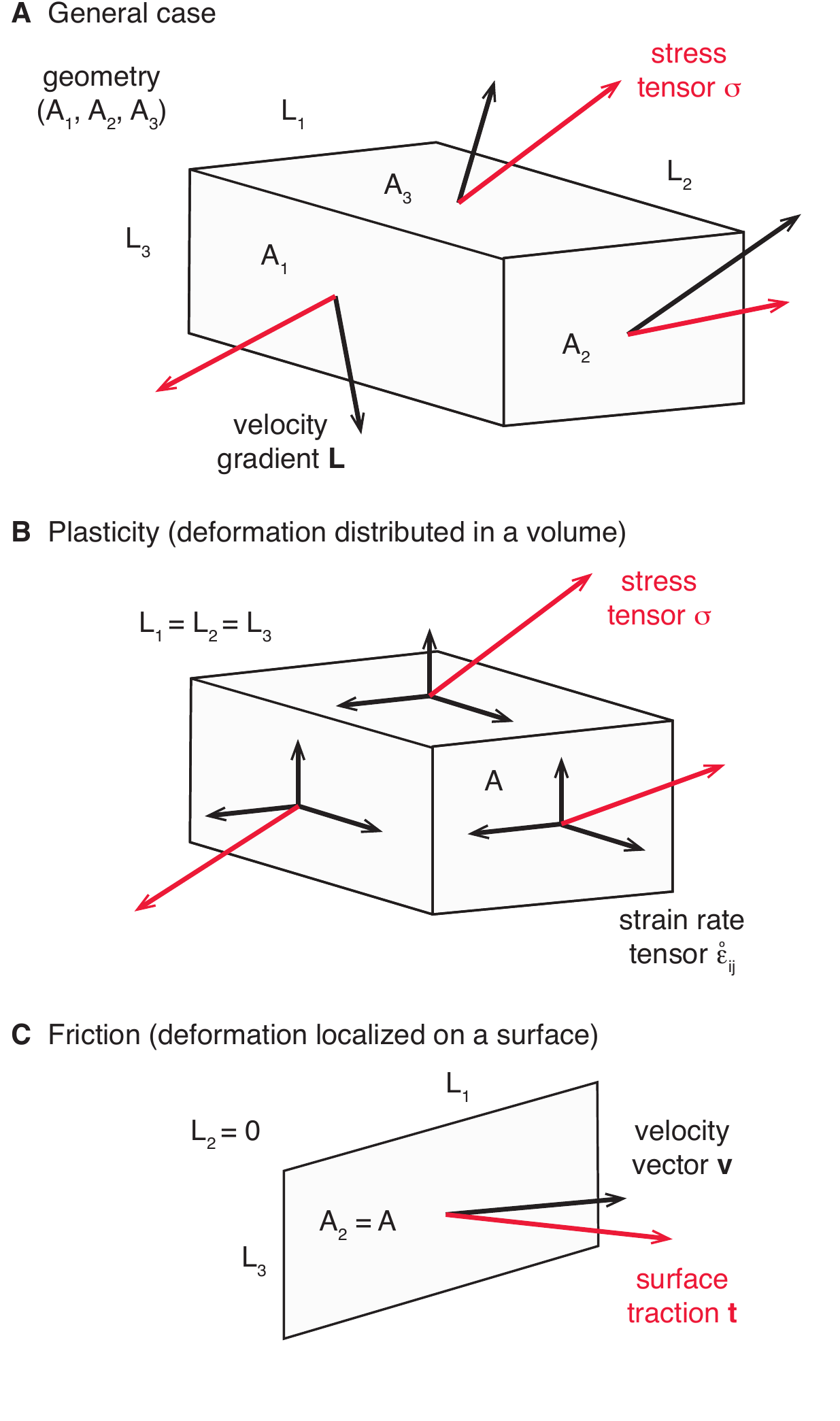}
\caption{A constitutive framework for plastic flow that incorporates the role of geometry. A) The spatial velocity gradient $\textbf{L}$ describes how the plastic velocity $\textbf{v}$ changes spatially. The velocity gradient is controlled by the stress tensor $\boldsymbol{\sigma}$ and the geometry of internal deformation, here modeled as a cuboid of dimension $L_1$, $L_2$, and $L_3$ with face areas $A_1$, $A_2$, and $A_3$. B) The constitutive laws of viscoplasticity, corresponding to a stress-strain-rate relationship, are obtained for $A_1=A_2=A_3$. C) The constitutive laws for kinematic friction is obtained when the deformed region collapses from a volume to a surface, here with $L_2=0$, leading to a velocity-traction relationship.}
\label{fig:unified-end-member}
\end{figure}

The unified kinematic description of distributed to localized deformation described in Section~\ref{sec:kinematic} and the similarities between the power-law viscoplastic laws and rate-and-state friction discussed in Section~\ref{sec:rheology} hint at the existence of a constitutive framework that would simplify to plasticity in some situations and to friction in others. 

In this section, I describe a unifying constitutive framework where the continuum between plastic flow and friction is only controlled by geometrical factors. When friction is interpreted as localized plastic flow, the constitutive relationship is anisotropic. Therefore, I introduce a heuristic that controls plastic anisotropy based on the geometry of internal deformation within a representative volume element. As internal tractions are modulated by the area on which they are applied, I consider the ratio of internal surface areas
\begin{equation}\label{eqn:tensor-state-variable}
\begin{aligned}
\textbf{A}&=\frac{1}{\underset{i}{\max}\,A_i}\left(\begin{matrix}
A_1 & 0 & 0 \\
0 & A_2 & 0 \\
0 & 0 & A_3 \\
\end{matrix}\right)~,
\end{aligned}
\end{equation}
as a dimensionless second-order tensor state variable of the representative volume element. In order to keep aspect ratios free variables, I consider a cuboid element of dimension $L_1$, $L_2$, and $L_3$ with face areas $A_1=L_2L_3$, $A_2=L_1L_3$, and $A_3=L_1L_2$ (Fig.~\ref{fig:unified-end-member}a). I define the weighted moment density tensor
\begin{equation}\label{eqn:weighted moment density tensor}
\begin{aligned}
\textbf{M}&=\textbf{A}\cdot\boldsymbol{\sigma}~.
\end{aligned}
\end{equation}
Each element of $\textbf{M}$ is proportional to a component of the force acting in opposite directions on faces $\textbf{e}_i$ and $-\textbf{e}_i$. I then define the weighted mean moment density
\begin{equation}
\sigma= \left. {\sum_i A_iM_{ii}} \middle/ {\sum_i A_i} \right.
\end{equation}
and the weighted mean moment density tensor as
\begin{equation}
\textbf{P}=\frac{1}{\sum_iA_i}\left(\begin{matrix}
A_1 & 0 & 0 \\
0 & A_2 & 0 \\
0 & 0 & A_3 \\
\end{matrix}\right)M_{kk}~.
\end{equation}
Most naturally, the definition of the deviatoric weighted moment density tensor follows as
\begin{equation}
\textbf{M}'=\textbf{M}-\textbf{P}~,
\end{equation}
with the apparent shear stress
\begin{equation}\label{eqn:apparent-stress}
\tau=||\textbf{M}'||~.
\end{equation}
Both plastic flow and kinematic friction can be formulated with a single constitutive law at the spatial scale of the representative volume element following
\begin{equation}\label{eqn:unified-rheology}
\textbf{L}=\epsilon_0\bigg(\frac{\tau}{\sigma}\bigg)^{n}\bigg(\frac{\theta}{\theta_0}\bigg)^m\frac{\textbf{M}'}{\tau}~,
\end{equation}
where $L_{ij}=v_{j,i}$ is the spatial anelastic velocity gradient, $\textbf{v}$ is the anelastic velocity field, and the anelastic strain rate is given by
\begin{equation}
\dot{\boldsymbol{\epsilon}}=\frac{1}{2}\left(\textbf{L}+\textbf{L}^T\right)~.
\end{equation}
The change of scale to the dimension of internal deformation requires the condition number of the state variable tensor
\begin{equation}\label{eqn:condition-number}
\kappa(\textbf{A})= \left. {\underset{i}{\max}\,L_i} \middle/ {\underset{i}{\min}\,L_i} \right.~,
\end{equation}
that quantifies the degree of localization within the representative volume element.

As an example, using the framework of equations~(\ref{eqn:tensor-state-variable}-\ref{eqn:condition-number}), I consider two end-member cases. First, I assume that the cuboid is a cube (Fig.~\ref{fig:unified-end-member}b), such as $L_1=L_2=L_3$ and $A_1=A_2=A_3=A$. The state variable becomes $\textbf{A}=\textbf{I}$ and the weighted moment density tensor $\textbf{M}=\boldsymbol{\sigma}$. The weighted mean moment density reduces to $\sigma=\frac{1}{3}\sigma_{kk}$ and the weighted mean moment density tensor simplifies to
\begin{equation}
\textbf{P}=\frac{1}{3}\sigma_{kk}\,\textbf{I}~,
\end{equation}
i.e., the isotropic pressure tensor. Finally, the deviatoric moment tensor reduces to
\begin{equation}\label{eqn:simplification-deviatoric-stress}
\textbf{M}'=\boldsymbol{\sigma}'~,
\end{equation}
i.e., the deviatoric stress tensor, which plays a central role in the theory of viscoplasticity. With $L_1=L_2=L_3$, the proposed constitutive law (\ref{eqn:unified-rheology}) reduces to the typical viscoplastic rheology
\begin{equation}
\dot{\boldsymbol{\epsilon}}=\dot{\epsilon}_0\left(\frac{\tau}{p}\right)^{n-1}\left(\frac{\theta}{\theta_0}\right)^m\boldsymbol{\sigma}'~,
\end{equation}
as the symmetry of the deviatoric stress $\boldsymbol{\sigma}'$ implies $\dot{\boldsymbol{\epsilon}}=\textbf{L}$ in this case. 

Now I consider a second case where the cuboid flattens to a surface (Fig.~\ref{fig:unified-end-member}c) with $L_2=0$, leading to $A_1=A_3=0$ and $A_2=A$. In practice, $L_2$ may be finite, but I consider the limit case. The state variable simplifies to 
\begin{equation}
\textbf{A}=\left(\begin{matrix}
0 & 0 & 0 \\
0 & 1 & 0 \\
0 & 0 & 0 \\
\end{matrix}\right)
\end{equation}
and the weighted moment density tensor follows as
\begin{equation}
\textbf{M}=\left(\begin{matrix}
0 & 0 & 0 \\
\sigma_{21} & \sigma_{22} & \sigma_{23} \\
0 & 0 & 0 \\
\end{matrix}\right)~,
\end{equation}
i.e., only the components corresponding to the traction vector $\textbf{t}^{(\textbf{e}_2)}$ are non zero and the tensor is not symmetric. The weighted mean moment density reduces to $\sigma=\sigma_{22}=\sigma_n$, where $\sigma_n$ is the normal stress to the surface $A_2$. The weighted mean moment density tensor becomes
\begin{equation}
\textbf{P}=\left(\begin{matrix}
0 & 0 & 0 \\
0 & \sigma_{22} & 0 \\
0 & 0 & 0 \\
\end{matrix}\right)~,
\end{equation}
where $\sigma_{22}$ is the normal stress $\sigma_n$ in this case. Finally, the deviatoric moment density tensor reads
\begin{equation}
\textbf{M}'=\left(\begin{matrix}
0 & 0 & 0 \\
\sigma_{21} & 0 & \sigma_{23} \\
0 & 0 & 0 \\
\end{matrix}\right)~.
\end{equation}
The norm of the deviatoric moment density tensor, $\tau=\sqrt{\sigma_{21}^2+\sigma_{23}^2}$, corresponds to the norm of the shear tractions that are parallel to the surface $A_2$, as is required to describe the local frictional resistance. The deviatoric moment density tensor is not symmetric either, allowing for simple shear. For the case where the cuboid region collapses to a surface with $L_2=0$, only two nontrivial components of the constitutive relationship remain
\begin{equation}\label{eqn:simplification-traction-velocity}
\begin{aligned}
v_{1,2}&=\epsilon_0\left(\frac{\tau}{\sigma_n}\right)^{n}\left(\frac{\theta}{\theta_0}\right)^m\cos\alpha \\
v_{3,2}&=\epsilon_0\left(\frac{\tau}{\sigma_n}\right)^{n}\left(\frac{\theta}{\theta_0}\right)^m\sin\alpha
\end{aligned}
\end{equation}
that correspond to the strike-slip and dip-slip component of the velocity gradient. Here, $\alpha$ is the rake angle, with $\cos\alpha=\sigma_{21}/\tau$ and $\sin\alpha=\sigma_{23}/\tau$. The spatial velocity gradient is not symmetric, creating simple shear, compatible with fault slip. Both $v_{1,2}$ and $v_{3,2}$ are finite because the internal deformation is averaged over the dimension of the representative volume element. The internal deformation is given by $\kappa(\textbf{A})\,\textbf{L}$, which approaches infinity with increasing localization. As described in Section~\ref{sec:rheology}, the formulation (\ref{eqn:simplification-traction-velocity}) corresponds to the multiplicative form of rate-and-state friction in three dimensions. 

In the proposed constitutive framework, whether the deformation simplifies to plasticity (distributed) or friction (localized) depends on a pair of geometrical factors in three dimensions, either the pair $L_1/L_3$ and $L_2/L_3$, the pair $L_1/L_2$ and $L_3/L_2$, or the pair $L_2/L_1$ and $L_3/L_1$. In two dimensions, if $L_1$ is infinite, the behavior will be controlled by the aspect ratio $L_2/L_3$.

The comparison of these two end-members, one corresponding to the distributed macroscopic stress and the other to the tractions on a surface, shows that friction can emerge from anisotropic plasticity in the special case of large aspect ratios in the internal geometry of the representative element.

\section{Conclusions}

The kinematics of plastic deformation in a continuum can incorporate all degrees of localization, from distributed to localized with a single mathematical expression. The constitutive law for kinematic friction closely resembles the power laws used in viscoplastic theories, in particular when the multiplicative form of rate-and-state friction is employed. 

A unifying constitutive framework can be identified, in which viscoplasticity and kinematic friction represent end-members of the rheological behavior for distributed and localized deformation, respectively. The behavior is controlled by the aspect ratio of the deformation region. It is also likely that different microphysical processes become activated depending of the aspect ratio. 

While I hope that this discussion will provide novel insight into the place of friction in the general framework of anelastic deformation, I have left out some important issues such as the physical origin of strain localization, in particular, what controls the thickness of a shear zone or fault gouge, and the true nature of the state variable used in (\ref{eqn:power-law-friction-velocity}) and (\ref{eqn:power-law-friction}), which is left to future work.

\section*{Acknowledgments}

This work comprises Earth Observatory of Singapore contribution no. 174. This research is supported by the National Research Foundation of Singapore under the NRF Fellowship scheme (National Research Fellow Awards No. NRF-NRFF2013-04) and by the Earth Observatory of Singapore, the National Research Foundation, and the Singapore Ministry of Education under the Research Centres of Excellence initiative.

\section*{References}

\bibliographystyle{unsrt}

\end{document}